\documentclass[11pt]{article}

\usepackage{colortbl}
\usepackage{amssymb}
\usepackage{color,graphicx}
\usepackage{hyperref}
\newcommand\fnurl[2]{%
\href{#1}{#2}\footnote{\url{#1}}%
}

\begin{document}




\title{Achieving Ethical Algorithmic Behaviour in the Internet-of-Things: a Review}

\author{Seng W. Loke\\School of IT, Deakin University, Geelong, Australia }

\maketitle

\begin{abstract}
The Internet-of-Things is emerging as a vast inter-connected space of devices and things surrounding people, many of which are increasingly capable of autonomous action, from automatically sending data to cloud servers for analysis, changing the behaviour of smart objects,  to changing the physical environment. A wide range of ethical concerns has arisen in their usage and development in recent years. Such concerns are exacerbated by the increasing autonomy given to connected things. This paper reviews, via examples, the landscape of   ethical issues, and some recent approaches to address these issues, concerning connected things behaving autonomously, as part of the Internet-of-Things.  We consider ethical issues in relation to device operations and accompanying algorithms.  Examples of concerns   include unsecured consumer devices,   data collection with health related Internet-of-Things, hackable vehicles and behaviour of autonomous vehicles in dilemma situations, accountability with Internet-of-Things systems, algorithmic bias, uncontrolled cooperation among things, and automation affecting user choice and control.  Current ideas  towards addressing a range of ethical concerns are reviewed and compared, including programming ethical behaviour, whitebox algorithms, blackbox validation, algorithmic social contracts, enveloping IoT systems, and guidelines and code of ethics for IoT developers - a suggestion from the analysis is that a  multi-pronged approach could be useful, based on the context of operation and deployment.

\end{abstract}








\section{Introduction} 
The Internet-of-Things (or IoT, for short) involves devices or things connected to the Internet or with networking capability. 
This includes Internet devices such as smartphones, smartwatches,  smart TVs, smart appliances, smart cars, smart drones, as well as everyday objects with Bluetooth, 3G/4G, and WiFi capabilities.
Specialised IoT protocols such as NB-IoT, Sigfox and LoRAWAN provide new connectivity options for  the IoT.\footnote{https://www.rs-online.com/designspark/eleven-internet-of-things-iot-protocols-you-need-to-know-about} 

{ Apart from  industrial IoT systems}, what is beginning to emerge is the notion of everyday objects with

\begin{itemize}
\item
  Internet or network connectivity (e.g., WiFi or Bluetooth enabled),
\item
  sensors (e.g., think of the sensors in the smartphone but also in a
  fork to detect its movement and how fast people eat~\cite{cite-key,Kadomura:2013:SFE:2468356.2468634}\footnote{See https://www.hapi.com/product/hapifork}), 
\item
  computational ability (e.g., with embedded AI~\cite{6766209} and cooperation protocols), { and
  \item actuators, or the ability to affect the physical world, including taking action autonomously.} 
\end{itemize}
{
The above highlights only some aspects of the IoT -  an extensive  discussion on the definition of the Internet of Things is in~\cite{iotdefn}.
}

There are also new home appliances like
\fnurl{https://developer.amazon.com/alexa}{Amazon Alexa} and
\fnurl{https://madeby.google.com/home/}{Google Home}, which have emerged
with Internet connectivity as central to their functioning, 
{ and often, they can be used to control other devices in the home}.
When things are not only Internet-connected but addressable via Web
links or URLs (the Uniform Resource Locators), and communicate via Web
protocols (e.g., using the Hypertext Transfer Protocol (HTTP)) the
so-called \fnurl{https://www.w3.org/WoT/}{Web of Things} emerges.

With increasing autonomy (fuelled by developments in Artificial Intelligence (AI)) and connectivity (fuelled by developments in wireless networking), there are a number of implications:

\begin{itemize}
  
  \item greater cooperation among IoT devices can now happen - devices that were previously not connected could now not only communicate { (provided time and resource constraints allow)} but carry out cooperative behaviours. In fact, the work in~\cite{7819416} envisions universal machine-to-machine collaboration across manufacturers and industries by the year 2025, { though this can be restricted due to proprietary data}; having a cooperation layer above the networking layer is an important development -  the social IoT has been widely discussed~\cite{Tripathy2016,doi:10.1155/2016/8178417,DBLP:journals/corr/LinD17,7397053};

\item
  network effects emerge: the value of a network is dependent on the
  size of the network; the greater the size of the network, the greater
  the value of joining or connecting to the network, so that device manufacturers could tend to favour cooperative IoT (e.g., see the
  \fnurl{https://www.technologyreview.com/s/527361/the-economics-of-the-internet-of-things/}{Economics
  of the Internet of Things}); a device that can cooperate with more devices could have greater value, compared to ones that cooperate with only a few - { such cooperation among devices can be triggered by users directly or indirectly (if decided by a device autonomously), { with consequent impact on communication latency and delay}};
\item
  devices which are connected to the Internet are controllable via the
  Internet, which means they are also vulnerable to (remote) hacking, in the same way that a computer being on the Internet can be hacked; 
\item
  sensors on such IoT devices gather significant amounts of data and,
  being Internet-enabled, such data are typically uploaded { to a server}
  (or to a Cloud computing server somewhere); potentially, such data can
  cause issues with people who are privacy-conscious (e.g., data from an  
  Internet connected light bulb could indicate when someone is home and not home); an often linked topic to the IoT is the notion of data
  analytics due to the need to process and analyze data from such
  sensing devices;
\item
IoT  devices might be deployed over a long time (e.g., embedded in a building or be part of urban street lighting) so that they need to be upgraded (or their software upgraded) over the Internet as improvements are made, errors are found and fixed, and as security vulnerabilities are discovered and patched;
\item
  non-tech savvy users might find working with Internet-connected
  devices challenging (e.g., set up and maintenance, and be unaware of security or privacy 
  effects of devices), { and users might feel a lost of control}; and
  \item
  computation on such devices suggests greater { autonomy} and more complex decision-making is possible { (and devices with spare capacity can also be used to supplement other devices)}; in fact, autonomous behaviour in smart things are not new' { smart things  detecting sensor-based context and responding autonomously (using approaches ranging from simple Even-Condition-Action rule-based  to sophisticated reasoning in agent-based approaches) have been explored extensively in context-aware computing~\cite{Loke:2006:CPS:1214754,Cristea2013}.}
  \end{itemize}

From the above, one can see that the IoT offers tremendous opportunity, 
 but also raises a range of ethical concerns. Prominent Computer Scientists have noted the need for ethical policies to guide IoT governance, in the areas of privacy rights, accountability for autonomous systems, and  promoting the ethical use of technologies~\cite{Berman:2017:SEB:3042068.3036698}.

  This paper aims to review the landscape of ethical concerns and issues that have arisen and which could, in future, arise with the Internet-of-Things, focusing on device operations and accompanying algorithms, especially as algorithms are increasingly  embedded in, and run on, IoT devices  enabling devices to take action with increasing autonomy; the paper also  reviews current ideas for addressing these concerns. 
 Based on the review, it is suggested that a multi-pronged approach can be useful for achieving ethical algorithmic behaviour in connected things.
 
 \subsection{Scope and Context}
    
    There have been much recent thinking on how Artificial Intelligence (AI) technologies can be integrated with IoT, from applying AI algorithms to learn from IoT data, multiagent views of IoT, to connected robots~\cite{6766209, 7980111}.\footnote{https://emerj.com/ai-sector-overviews/artificial-intelligence-plus-the-internet-of-things-iot-3-examples-worth-learning-from/}
As Artificial Intelligence (AI) capabilities become embedded into IoT devices, the  devices gain greater autonomy and decision-making capabilities, automating a wider range of tasks, so that some things can be described as ``robotic''.
For example, we can imagine a bookshelf that one can talk to and which can serve us, relocating and reorganising books at our command, or a library where storage and retrieval of (physical) books { is} automated,  or think of a standing lamp that follows and tracks the user as the user moves around on the sofa or in the room  - a question is whether  these libraries and the standing lamp can be considered ``robots''. 
Also, autonomous connected vehicles~\cite{driverlessbook},  with Internet-enabled networking and vehicle-to-vehicle connectivity, have also captured the world's imagination and have enjoyed tremendous development in recent years.
The discussion in this paper, hence, includes  robots, AI as used with IoT, and autonomous vehicles, under a broad definition of IoT. The link between IoT and robotics has also been made in~\cite{doi:10.1177/1729881418759424,7805273}, yielding the notion of the Internet of Robotic Things.

   Ethics in AI has been extensively discussed elsewhere (e.g., ~\cite{ijcai2018-779} and ethical AI principles\footnote{https://futureoflife.org/ai-principles/}), and indeed, the integration of AI technologies into the IoT,  as mentioned above, calls for  ethical AI principles to be considered in the development of such IoT  systems. Hence, this paper reviews work not only in ethical issues in IoT but also includes a review of work on ethics in AI and robotics in the context of IoT.

 While this paper reviews technical approaches  mainly from computing  - the issues are often interdisciplinary, at the intersection of computing technology, philosophy (namely, ethics), law and governance (in so far as policies are involved),
 as well as diverse application domains (e.g., health, transport, business, defence, law, and others) where IoT has been applied.  
  Moreover, while security and data privacy are key concerns in relation to things behaving ethically, the concerns on ethical behaviour go beyond security and privacy.
 The field is also continually growing in recent years as ethical issues for IoT are highlighted in mass media and with much research in the area (examples highlighted in the following sections), and hence, the paper does not seek to completely cover all recent work, but can only  provide a comprehensive snapshot and introduction to the area, while highlighting potential approaches to  the issues.

 The seminal review on ethics of computing~\cite{Stahl:2016:ECS:2891449.2871196} lays out five aspects of each paper reviewed: ethical theory that aids interpreting the issue, the methodology applied, the technology context, and contributions and recommendations. Different from~\cite{Stahl:2016:ECS:2891449.2871196},  this paper paper focuses on the ethical issues in IoT work, but, indeed, these aspects have informed the reading of work in this area at the junction of the IoT and ethics.  This paper  touches on a range of ethical issues noted in the paper, namely, agency, autonomy, consent, health,   privacy, professionalism, and trust in our discussions. For example, we discuss issues of user choice and consent in IoT devices, autonomy of things in their function, consider health IoT issues, security, trust and privacy of IoT devices, and code of ethics for IoT developers. We do not discuss ethical issues in relation to inclusion and the digital divide but retain a technical focus in this paper.
 
 The survey on foundational ethical issues in IoT~\cite{ALLHOFF201855} focused on informed consent, privacy, security, physical safety, and trust and noted, importantly, that these are inter-related in IoT. This paper also discusses  a range of these issues but we also consider examples and solutions (many originally from outside typical IoT research areas)  to achieving ethical IoT systems.

\subsection{Organization}
 The rest of the paper is organised as follows. { To introduce readers to ethical issues in IoT, the next section first  discusses, via examples, ethical concerns with IoT. Then, the following section examines ideas which have been proposed to address these concerns, and notes the need for a multi-pronged approach.}
 The final section concludes with future work.




\section{Ethical Concerns and Issues}

This section reviews ethical concerns and issues with IoT devices and systems, via examples in multiple application domains, including the need for consumer IoT devices to employ adequate security measures, ethical data handling by health related IoT systems, right  behaviour of autonomous vehicles in normal usage and dilemma situations, usage concerns with connected robots and ethical robot behaviour, algorithmic bias that could be embedded into IoT systems,    
 right behaviour when IoT devices cooperate, and user choice restrictions or lost of control  with automated IoT systems.
 { Below, the unit of analysis is either an
 individual IoT device or a collection (or system)  of such IoT devices (the size of which depends on the application).}

\subsection{Unsecured Consumer IoT Devices}


{
The security and data privacy issues in IoT are well surveyed and have been discussed extensively, e.g., in \cite{SHA2018326,GE201712,SICARI2015146,MALINA201683,Atwady:2017:SAT:3102304.3102312,RIAHISFAR2018118,ALABA201710,TRNKA,DIAZ201699,JAYARAMAN2017540}. 
The contents of the surveys are not repeated here but some examples of issues with unsecured IoT devices are highlighted below.
}

There are  IoT devices which may have been shipped without
encryption  (with lower computational
power devices which are not capable of encrypted communications). A study
by HP\footnote{https://community.softwaregrp.com/t5/Protect-Your-Assets/HP-Study-Reveals-70-Percent-of-Internet-of-Things-Devices/ba-p/220516\#.WiY9QmLZXDv} suggested 70 percent of IoT devices use unencrypted
communications. However, it must be noted that cheaper does not necessarily mean less secure as cost depends on a range of factors beyond security capability.

A Samsung TV was  said to listen in on living room conversations
as it picks up commands via voice recognition. The company has since clarified
that it does not record correctly conversations arbitrarily.\footnote{http://abcnews.go.com/Technology/samsung-clarifies-privacy-policy-smart-tv-hear/story?id=28861189}  However,
it does a raise a concern with devices in the same category as voice-activated or conversational devices, whether they do record conversations.

In an experiment\footnote{https://www.timesofisrael.com/israeli-hackers-show-light-bulbs-can-take-down-the-internet/}  at Israel's Weizmann Institute of Science, researchers managed to
fly a drone within 100 metres of a building and remotely infect light
bulbs in the building by exploiting a weakness in the ZigBee Light Link
protocol, used for connecting to the bulbs. The infected bulbs were then
remotely controlled via the drone and made to flash `SOS'.

A report
on the Wi-Fi enabled Barbie doll\footnote{https://www.theguardian.com/technology/2015/nov/26/hackers-can-hijack-wi-fi-hello-barbie-to-spy-on-your-children}
 noted that they can be hacked and
turned into surveillance devices. This was then followed by a FBI advisory note on
IoT toys,\footnote{https://www.ic3.gov/media/2017/170717.aspx} about possible risk to private information disclosure.
A 11-year old managed to hack
into a Teddy Bear via Bluetooth.\footnote{https://securityintelligence.com/news/with-teddy-bear-bluetooth-hack-11-year-old-proves-iot-security-is-no-childs-play/} 
And the ubiquitous IoT cameras have
certainly not
been free from hacking.\footnote{http://www.zdnet.com/article/175000-iot-cameras-can-be-remotely-hacked-thanks-to-flaw-says-security-researcher/}

There are many
other examples of IoT devices getting hacked.\footnote{https://www.wired.com/2015/12/2015-the-year-the-internet-of-things-got-hacked/} 
As research shows,\footnote{https://arxiv.org/pdf/1705.06805.pdf} someone can still infer
people's private in-home activities by monitoring and analysing network
traffic rates and packet headers, even when devices use encrypted
communications.

{
The above are only a few of many examples and has implications for
developers of IoT devices which must incorporate security features, for 
policy-makers, for cyber security experts, as well as for users who would
need to be aware of potential risks. 

Recent surveys also highlighted privacy and managerial issues with the IoT~\cite{WEINBERG2015615}.
From the Australian privacy policy perspective~\cite{CARON20164}, after a review of the four issues of (1) IoT based surveillance, 
(2) data generation and use, (3) inadequate authentication and (4) information security risks, the conclusion is that the Australian Privacy Principle is inadequate for protecting the individual privacy of data collected using IoT devices, and that given the global reach of IoT devices, privacy protection legislation is required across international borders.
Weber~\cite{WEBER2015618} calls for new legal approaches to data privacy   in the IoT context, from the European perspective, based on improved transparency and data
minimization principles. The recent European regulation, the General Data Protection Regulation (GDPR)\footnote{https://eugdpr.org/} is a law aimed at providing people with greater control of their data, and has implications and challenges for IoT systems, with requirements on systems such as privacy-by-design, the right to be forgotten or data erasure, the need for clarity in requesting consent,  and data portability where users have the right to receive their own data, as discussed in~\cite{8453643}. Companies are already coming on board with tools to support GDPR requirements.\footnote{For example, see Microsoft's tools: https://www.microsoft.com/en-au/trust-center/privacy/gdpr-overview and Google: https://cloud.google.com/security/compliance/gdpr/}

A recent 
\fnurl{https://citp.princeton.edu/event/conference-internet-of-things/}{workshop on privacy and security policies for IoT at Princeton University} has
raised a range of  issues, suggesting a still on-going
area of research  at the intersection of IT, ethics, governance and law.
Cyber physical systems security  is discussed extensively elsewhere~\cite{7924372}.
}

{
Security also impacts on usability since additional measures might be taken to improve security, for example, when users have to reset passwords before being allowed to use a device, the use of multi-factor authentication schemes, and configuration set up to improve security during use, all of which could reduce usability; the work in~\cite{Dutta2017} highlights the need to consider the usability impact of IoT security features at design time.
}


\subsection{Ethical Issues with Health Related IoT}
IoT medical devices are playing an increasingly critical role in human life, but as far back as 2008, implantable
 defibrillators have been known to be `hackable'~\cite{4531149},
allowing communications to them to be intercepted.

Apart from the security of  IoT devices, in~\cite{Mittelstadt2017}, a range of ethical issues with the use of IoT in health were surveyed, including:


\begin{itemize}
\item {\em personal privacy}:  this relates not just to privacy of collected data, but the notion that a person has the freedom not to be observed or to have his/her own personal space; the use of smart space monitoring (e.g., a smart home or in public spaces such as aged care facilities) of its inhabitants raises the concern of continual observation of people, even if it is for their own good { - being able to monitor individuals or groups can be substantially beneficial but presents issues of  privacy and access};

\item {\em informational privacy}: this relates to one's ability to control one's own health data - it is not uncommon for organizations to ask consumers for private data with the promise that the data will not be misused - in fact, privacy laws could prohibit use of the data beyond its intended context - the issues are myriad (e.g., see~\cite{Chamberlain2017}), including how one can access data collected by an IoT device but now possibly owned by the company, how much an insurance company could demand of user health data,\footnote{https://www.iothub.com.au/news/intel-brings-iot-to-health-insurance-411714} how one can share data in a controlled manner, how one can prove the veracity of personal health data, and how users can understand the privacy-utility tradeoffs when using an IoT device;

\item {\em risk of non-professional care}: the notion of self health-monitoring and self-care as facilitated by health IoT devices  can provide a false optimism, limiting a patient's condition to a narrow range of device-measurable conditions; confidence in non-professional carers armed with IoT devices might be misplaced.

\end{itemize}

{
The above issues relate mainly to health IoT devices but the data privacy issues apply to other Internet connected devices in general~\cite{RePEc:scm:usvaep:v:13:y:2013:i:2(18):p:210-216}.\footnote{See also http://arno.uvt.nl/show.cgi?fid=144871} Approaches to data privacy range from privacy-by-design, recent blockchain-based data management and control (e.g.,~\cite{Ali:2017:IDP:3131542.3131563,8370027, Griggs2018,REYNA2018173,8436081}), to regulatory frameworks that aim to provide greater control over data to users as reviewed earlier, e.g. in~\cite{8453643,DIAZ201699,SICARI2015146,JAYARAMAN2017540}. 
{ There are also issues related to how health data should or should not be used - for example, what companies are allowed to use the health data for (e.g., whether an individual could be denied insurance based on health data, or denied access to treatment based on lifestyle monitoring).}

In relation to IoT in sports to help sports training and fitness tracking, incorporated in an artificial personal trainer, there are numerous technical  challenges~\cite{FISTER2015178}, including { generating and adapting plans for the person, measuring the person's readiness, personal  data management, as well as validation and verification of fitness data}.  One could think of issues and liability with errors in measurement or an athlete being endangered or subsequently even injured over time by erroneous advice due to incorrect measurements, or issues arising from following the advice of an AI trainer or such devices being hacked. In any case, there are already several wearable personal trainers on the market\footnote{E.g., see https://welcome.moov.cc/ and https://vitrainer.com/pages/vi-sense-audio-trainer} 
 which come with appropriate precautions and disclaimers for users\footnote{See https://welcome.moov.cc/terms/ and https://vitrainer.com/pages/terms-and-conditions} and privacy policies.
 }

 \subsection{Hackable Vehicles and the Moral Dilemma for Autonomous
Vehicles (AVs)}\label{moral-dilemma-for-autonomous-vehicles}

Cars with computers are not unhackable, an example is the Jeep which was hacked while on the road,\footnote{https://www.wired.com/2015/07/hackers-remotely-kill-jeep-highway/}
and made to be remotely controllable. With many vehicles having Internet connectivity, their hackability is now public knowledge.\footnote{For example, see the online book on car hacking, http://opengarages.org/handbook/}
Similar security issues of encrypting communications with the vehicle,  securing the vehicle system arises, and managing data collected by the vehicle, arise as in other IoT systems - security issues for autonomous vehicles are discussed elsewhere~\cite{Lima:2016:TSS:2994487.2994489,autosar}. Given the wide range of data collected about the vehicle, from telemetry data to location data, as well as logs of user interaction with the vehicle computer, privacy management of vehicular data { is an issue}~\cite{7856736}.

Recent developments in autonomous vehicles have provided tremendous promise for reducing road injuries and deaths due to human error, as well as the potential to start a  `new' industry, with many countries around the world working on autonomous vehicle projects,\footnote{https://avsincities.bloomberg.org/global-atlas}  with subsequent impact on the design of cities.\footnote{See http://www.nlc.org/AVPolicy and https://www.wired.com/2016/10/heres-self-driving-cars-will-transform-city/}
 However, as autonomous vehicles function in a socio-technical environment, there could be
decisions they need to make, which involves moral reasoning as discussed in~\cite{Bonnefon1573}.\footnote{See also this TED talk by \fnurl{https://rahwan.me/}{Professor Iyad Rahwan
from MIT}}


Essentially, the moral dilemma of autonomous vehicles is similar to the trolley problem in
philosophy{\footnote{See https://fee.org/articles/the-trolley-problem-and-self-driving-cars/}} - suppose an autonomous vehicle is about to hit someone in
its way but the only way to avoid this is to swerve to the right or
left, but will kill some  pedestrians while doing so - { or should it opt to protect the occupants of the vehicle in preference to external individuals.} Either way,
someone will be killed, what should the autonomous vehicle do?\footnote{ Such dilemma situations can occur in other smart things scenarios - { e.g., consider this original example:} in a fire situation, a smart door can decide to open to let someone through but, at the same time, would allow  smoke to pass in to possibly harm others, or a smart thing can choose to transmit messages about a lost dementia person frequently to allow finer-grained tracking for a short time, but risk  the battery running out sooner (and so losing the person, if not found in time), or transmit less frequently allowing longer operating time but coarse-grained location data.}
While there may be no clear-cut answer for the question, it is important
to note the ethical issue raised -  potential approaches to the problem are discussed later.
While AVs will help many people, there are issues about what the AVs
will do in situations where trade-offs are required. A utilitarian
approach might be to choose the decision which { potentially kills fewer people}. A
virtue ethics approach will not approve of that way of reasoning. A
deontological or virtuous approach might decide `not to kill' whatever the
situation, in which case, the situation cannot be helped.
One could also argue that such situations are unlikely to arise, but
there is also a small possibility that it could arise, and perhaps in many
different ways. Imagine an AV in a busy urban area receiving an
instruction to speed up due to a heart attack just happening in its
passenger, but this puts other pedestrians and road users at greater
risk - should the AV speed up? { However, one could  also note that sensors in the vehicle could detect that the passenger has a heart attack and report this to traffic management to have a path cleared, and so, speeding up may not be an issue - connectivity, hence, can help the situation rather than increase risk, while the ethics in the decision-making  remains challenging.} 


Ethical guidelines regulating the use and development of AVs are being 
developed - Germany was the first country to provide ethical guidelines for autonomous vehicles via the Ethics Commission on Connected and Automated Driving.\footnote{See the report at  https://www.bmvi.de/SharedDocs/EN/Documents/G/ethic-commission-report.pdf?\_\_blob=publicationFile}  The guidelines include an admission that autonomous driving cannot be completely safe: ``{\em ... at the level of what is technologically
possible today, and given the realities of heterogeneous and non-interlinked road
traffic, it will not be possible to prevent accidents completely. This makes it essential that
decisions be taken when programming the software of highly and fully automated driving
systems.}''
As noted in~\cite{Lin2016}, for ``{ {\em future autonomous cars, crash-avoidance features alone won't be enough}}'', but
when crash is inevitable, there needs to be a crash-optimization strategy but that strategy should not aim only to minimise damage - since if that was the case, the vehicle might decide to crash into a cyclist wearing a helmet than a cyclist not wearing a helmet, hence, targeting people who chose to be safer - there is no clear resolution of this ethical issue, as yet.

There are also issues concerning
who will take responsibility when something bad happens in an autonomous vehicle - whether it would be the passengers, the manufacturer or middlemen.
The issue is complex in mixed scenarios where human drivers and autonomous vehicles meet in an incident, and the fault lies in the human driver, but the autonomous vehicle was unable to react to the human error.

But assuming the success of autonomous vehicles to reduce road deaths and accidents, would it then be ethical to allow human drivers?
The work in~\cite{SPARROW2017206} goes as far as to suggest: {\em
``...making it illegal to manufacture vehicles that allow for manual driving beyond a certain date and/or making it illegal, while on a public road, to manually drive a vehicle that has the capacity for autonomous operations.}''
Appropriate policies for autonomous vehicles continues to be an open issue~\cite{Bagloee2016}.
{
Further approaches to ethical automated vehicles will be discussed in Section~III.
 }
\subsection{Roboethics}


Roboethics~\cite{Lin:2014:REE:2616404,robotethics2,spyros2016} is concerned with positive and negative aspects of robotics
technology in society, and explores issues concerning the ethical
design, development and use of robots.
While there are tremendous opportunities in
robotics, their widespread use also raises ethical concerns, and as the line between robots and autonomous IoT becomes blurred, the issues of ethics with robots are inherited by IoT.

\subsubsection{Robots Rights}\label{should-robots-have-rights}
There are some schools of
thought that have begun to ask the question of whether robots
(if capable of moral reasoning) should have rights~\cite{Gunkel2017}, and what level of autonomous decision-making would require robots to have rights, similar to how animals might have rights.
Indeed, the level of autonomy and sentience required of such machines before rights becomes an issue might still be far off. 
In fact, roboethics has largely been concerned with ethics that developers and
users of the technology need to consider.  
Below, we explore examples of ethical issues in robotics for surgery, personal assistance, and war.

\subsubsection{Robotic Surgery}\label{robotic-surgery}

Robots are capable of surgical operations as we have seen{, typically under the direction and control of a surgeon}. In 2000, the
U.S. Food and Drug Administration (FDA) approved the use of the Da Vinci
robotic surgical system for a surgeon to perform laparoscopic gall
bladder and reflux disease surgery. Robotic
surgery devices continue to be developed,\footnote{https://spectrum.ieee.org/robotics/medical-robots/would-you-trust-a-robot-surgeon-to-operate-on-you} and some make decisions
autonomously during surgeries, e.g., to automatically position a frame
for the surgeon's tools, where to cut bones, and delivering radiation for
tumours.  If costs of robotic surgery could be reduced, complex surgery could 
perhaps be made available to more people in third-world and developing countries.
As they get better, and can provide surgical help at lower costs, what is problematic is then not their use but denying people their use.



{
But an issue emerges when something goes wrong and the question of accountability and liability arises regarding the patient's injury. While one might not consider surgical robots as IoT devices, the issue of IoT devices making decisions that could
result directly in injury or harm, even if they were intentionally made to help humans raises similar concerns.
 
\subsubsection{Social and Assistive
Robots and Smart Things}\label{social-and-assistive-robots}

Social robots might play the role of avatars (remotely representing
someone), social partners (accompanying someone at home), or cyborg
extensions (being linked to the human body in some way).
A robot capable of social interaction might be expected to express and
perceive emotions, converse with users, imitate users, establish social
connection with users via gesture, gaze or some form of natural
interaction modality, as well as perhaps present a distinctive
personality. {
While they can be useful, some concerns include:
\begin{itemize}
\item Social robots or IoT devices may be able to form bonds with humans, e.g., an elderly
person or a child. A range of questions arises such as whether such
 robots should be providing emotional support in place of
humans, if they can be designed to do so.
Another question is what psychological and
physical risk of humans forming such bonds with such devices or robots - when a user is
emotionally attached to a thing, a concern is what would happen if the thing is
damaged or no longer supported by the manufacturer, or what if such things can be
hacked to deceive the user. { This question can be considered for smart things which has  learnt  and adapted to the person's behaviour and not easily replaced.}

\item Such social robots or IoT devices can be designed to have authority 
 to provide reminders, therapy or rehabilitation to users.
Ethical issues can arise when harm or injury is caused due to
interaction with such robots. For example, death is caused from
medication taken at the wrong time due to a robot's reminder at the
wrong time due to a malfunction. A similar concern carries over to a smart pill bottle (an IoT device) intended to track when a person has and has not taken medication with an associated reminder system.
There is also a question of  harm being 
caused inadvertently, e.g., when an elderly person trips over a robot that
approached too suddenly, or a robot makes decisions on behalf of
its owner, without the owner's full consent or before the owner could
intervene.

\end{itemize}
It must be noted, though, that the concerns above relate to behavioural aspects of the devices, not so much to the connectivity that the devices might have.
}

Ethical guidelines regarding their development and use are required,
including training of users and care-givers, affordability of such
devices, and prevention of malpractice or misuse.

Ethical principles for socially assistive robots are outlined
\fnurl{https://robotics.usc.edu/publications/media/uploads/pubs/689.pdf}{here},
including

\begin{quote}{\em 
``The principles of beneficence and non-maleficence state that caregivers
should act in the best interests of the patient and should do nothing
rather than take any action that may harm a patient.''}
\end{quote}

A similar guideline informs socially assistive { smart things, not only robots - how smart things with intelligent and responsive behaviours and} robots could
be programmed to provably satisfy those principles is an open research
issue. It remains an open research issue how { smart things and robots} could learn human values and be
flexible enough to act in a context-aware manner. 
{ Issues specifically due to the fact that these devices might be connected are similar to other IoT devices, e.g., sensitive private data possibly shared beyond safe boundaries, and vulnerability to remote hacking -  perhaps made worse by their close interaction with and proximity to users.
}


\subsubsection{Robots in War}\label{robots-in-war}

{
 Robots can be
used to disarm explosive devices, 24/7 monitoring, or
for risky spying missions, and engage in battles in order save lives.
But there are already controversies surrounding the use of automated
drones (even if remotely human-piloted) in war~\cite{dronewar}. While human casualties
can be reduced, the notion of humans being out of the loop in
\emph{firing decisions} is somewhat controversial. AI also might not
have adequate discriminatory powers for its computer vision technology
to differentiate civilians from soldiers. While robots can reduce human lives lost at war, there is also the issue
that it could then lower barriers to entry and even `encourage' war,
or be misused by `tyrants'. 
{
There have been the call for a ban on autonomous weapons in an
open
letter signed by 116 founders of robotics and AI companies from 26
countries,\footnote{https://futureoflife.org/2017/08/20/killer-robots-worlds-top-ai-robotics-companies-urge-united-nations-ban-lethal-autonomous-weapons/}
 and the  \fnurl{http://www.stopkillerrobots.org/}{\em Campaign to
stop Killer Robots.}}  
Algorithmic behaviour can be employed in remotely controlled robots to help human operators, but remote controlled and autonomous robotic weapons, if allowed,  will need to be designed for responsibility, i.e., allow human oversight and control.\footnote{https://www.oxfordmartin.ox.ac.uk/downloads/briefings/Robo-Wars.pdf} Robot-on-robot warfare might still be legal and ethical.
}


 
\subsection{Algorithmic Bias and IoT}

We explore the notion of bias in algorithms in this section. The following types of concerns with algorithms were noted by~\cite{doi:10.1177/2053951716679679}: {\em inconclusive evidence} when algorithms make decisions based on  approximations, machine learning techniques, or statistical inference, {\em inscrutable evidence}  where the workings of the algorithm is not known, and {\em misguided evidence} when the inputs to the algorithm are inappropriate.
{ Some automated systems have behaviour which can be opaque, unregulated and could amplify biases~\cite{womd}, inequality~\cite{autoinequality}, and racism~\cite{aoo}.}

Note that while such bias are not specifically situated in IoT systems, and there are IoT systems which do not interact with humans directly,  such issues are relevant as} there are also IoT devices with Internet applications that often employ face recognition algorithms, voice recognition algorithms (e.g., Google Home and Amazon Echo) and aim to present users with a summary of recent news and product recommendations.

Algorithmic bias can arise in autonomous systems~\cite{Danks:2017:ABA:3171837.3171944}, and could arise in IoT devices as they become increasingly autonomous devices.
An IoT device that behaves using a machine learning algorithm, if trained on bias data could yield bias behaviour. 
With the increasing data-driven nature of IoT devices, a number of possible opportunities for discrimination can arise as noted in~\cite{regulatingiot} -
examples given include an IoT gaming console and neighbourhood advisor that advises avoiding certain areas. 
 Also, such algorithmic bias can be in machine learning algorithms used for autonomous vehicles, where  large volumes of data over time frames of minutes to days are analysed.

Even without using machine learning, it is not to difficult to think of devices that can exhibit  biased behaviour  -  consider a sensor  that is biased in the information it captures, intentionally or unintentionally, or a robot that greeds certain type of people. Such a robot might be programmed to randomly choose who it greets, but it may happen to appear to only greet certain individuals, and so, be perceived as bias. 

 \subsubsection{Racist Algorithms}


While the algorithms or their developers might not be intentionally racist,
as machine learning thrives on data they are trained on, bias can be
introduced, even unintentionally. Hence, an algorithm may not be built
to be intentionally racist but a failure of a face recognition algorithm 
on those with darker skin\footnote{For example,  see https://www.ted.com/talks/\-joy\_buolamwini\_how\_i\_m\_fighting\_bias\_in\_algorithms/\-discussion?curator=TechREDEF}
colour could raise concerns and cause a category of people to feel
discriminated against.  
A device that has been trained to work in a certain context might not work as expected in a different context - a type of  transfer bias - a simple example of a smart door trained to open based on recognizing  fair-skinned persons might not open for  dark-skinned persons. 

How IoT devices interface  with humans could be bias by design, even if not intentionally so, but simply due to inadequate consideration.



As reported in the Technology
Review on bias in natural language processing systems,\footnote{https://www.technologyreview.com/s/608619/ai-programs-are-learning-to-exclude-some-african-american-voices/} due to the use
of machine learning to learn how to recognise speech, there are issues
with minority population groups due to lack of training examples for the
machine learning algorithms:
{\em ``If there aren't enough examples of a particular accent or vernacular,
then these systems may simply fail to understand you.''}
The original intention and motive of developers could
be considered when judging algorithmic bias, and care is needed to
determine if and when bias does arise, even if not originally intended,
especially with machine learning on data.

\subsubsection{Other Algorithmic Bias}

We have looked at how algorithms might appear to be racially biased in
its inference, but there could be other bias in general. For example, in politics,
where the algorithm tends to favour a given political party more than
others, or in business, where a particular brand of goods is favoured
over others. And suppose an algorithm  used in recommending news
articles or products for you does so in a systematically bias manner - it could then have an influence in your
voting or buying behaviours.  An algorithm that provides possibly biased recommendations or news is an issue that has put
Facebook in the news, when it was said to
{\em  ``deliberately
suppressing conservative news from surfacing in its Trending Topics}''.\footnote{https://www.wired.com/2016/05/course-facebook-biased-thats-tech-works-today/}
To provide greater transparency, Facebook also begin to describe how it
recommends and filters news in the Trending Topics section,\footnote{http://fortune.com/2016/05/12/facebook-and-the-news/} perhaps in
being more open to the public. Other concerns are on how Twitter provides
algorithmically filtered news feeds to users.\footnote{http://fortune.com/2016/02/08/twitter-algorithm/}


{
But what if the agenda is a ``good'' one, e.g.  algorithms being informed
by a utilitarian mandate. But this raises the ethical question of whether  software should be
programmed to always benefit as many people as possible, even at the cost of a 
few - considering a hypothetical ``smart'' water rationing system in homes, where water is conserved for all at the sacrifice of some urgent uses.} {  Also, taking a broader sustainability view, IoT systems can help cities move towards smarter more energy-efficient homes, smarter waste management and smarter energy grids, helping to achieve sustainable development goals.\footnote{For example, see https://deepblue.lib.umich.edu/bitstream/handle/2027.42/\\136581/Zhang\_TheApplicationOfTheInternetOfThingsToEnhance\\UrbanSustainability.pdf?sequence=1\&isAllowed=y and \\http://www3.weforum.org/docs/IoTGuidelinesforSustainability.pdf} Another example is IoT-based infrastructure monitoring helping to reduce urban flooding.\footnote{https://www.weforum.org/agenda/2018/01/effect-technology-sustainability-sdgs-internet-things-iot/} However, how, in general,  automated IoT systems should balance priorities within an overall sustainability agenda, without bias towards or against any community groups, can be a consideration from the system design phase.
}


Moreover, there could be an issue with human values and bias being
essentially incorporated into algorithms { or into the design of IoT devices}. The so-called {\em value-laden
algorithms} by \cite{Kraemer2011} can be defined as follows:
{\em ``An algorithm comprises
an essential value-judgment if and only if, everything else being equal,
software designers who accept different value-judgments would have a
rational reason to design the algorithm differently (or choose different
algorithms for solving the same problem).''}
An example discussed is that of medical image
algorithms. It is noted that ``{\em medical image algorithms
should be designed such that they are more likely to produce false
positive rather than false negative results}.''
However, the increased number of false positives might 
lead to too many unnecessary   operations.
Also, this could cause alarming results for many who are then
suspected or diagnosed with diseases that they do not have, due to the
medical image algorithms conservatively highlighting what is possibly
not there. Hence, due to the need to be conservative and to avoid
missing a diagnosis, a developer of the algorithm could have made it
more pessimistic so that nothing is overlooked. Or consider an Internet camera to detect intruders which gives too many false positives, in trying to be ``overly protective''.

To be fair, algorithmic bias can arise due to the developers own values
or due to data used in training algorithms, or simply due to cases not
considered during design, and perhaps not due to malicious or
intentionally biased agendas. However, an issue is how to distinguish
between intentional (and malicious) and unintentional algorithmic bias.

 \subsection{Issues with Cooperative IoT}
 
 When IoT devices cooperate, a number of issues arise. For example, with autonomous vehicles, 
it is not only vehicle-to-vehicle cooperation, an autonomous vehicle could share roads with pedestrians, cyclists, and other human-driven vehicles,
and  would need to  reason about social situations,  perform social signalling  with people via messaging or physical signs, and work within rules and   norms for the road, which could prove to be a difficult problem.\footnote{https://spectrum.ieee.org/transportation/self-driving/the-big-problem-with-selfdriving-cars-is-people,http://urban-online.org/en/human-factors-in-traffic/index.html}  

Protection from false messages, and groups of vehicles that cooperate maliciously, are also concerns, looking forward. How will a vehicle know if a message to make way is authentic? What if vehicles take turns to dominate parking spaces or gang-up to misinform non-gang vehicles about where available parking might be? Or what if vehicles of a  given car manufacturer collude to disadvantage other brands of cars.

A similar issue arises with other IoT devices which must discern the truthfulness of messages they receive, and which, when cooperating, and exchanging data would need to follow policies for data exchanges.
Denial-of-Service attacks where a device receives too many spurious messages must be guarded against and IoT devices should not   spam other devices. The issues of trust with a large number of inter-connected devices has been explored, with a proposed trust model, in~\cite{DBLP:journals/corr/LinD17}.
 
 { With cooperation,  considerations of what data should be shared and how data is shared among cooperative IoT devices will be  important.  For example, if a group of vehicles share routing information in order to cooperate on routing to reduce congestion, as in ~\cite{DBLP:journals/tits/DesaiLDS13}, there is a need to ensure that such information is not stored or used beyond their original purpose. 
 }
 
{
\subsection{User Choice and Freedom}
Apart from transparency of operations, systems allowing adequate user choice is also important - freedom of action is an important property of  systems that are  respectful of the autonomy of users, or at least a user's direction is based on the user's own {\em ``active assessment of factual information}''~\cite{10.2307/20014493}.\footnote{http://www.ethics.org.au/on-ethics/blog/october-2016/ethics-explainer-autonomy} For example, a device can be programmed to collect data and manage data automatically (e.g., once a photo is taken by a device, it can be  automatically shared with a number of parties and stored), but people would like to be informed about what data is collected and how data is used.  Informing might not be adequate - a system could automatically inform a user that all photos on a smartphone will be copied to the cloud and will be categorised in a default manner on the phone, but the user might  want control over what categories to use and control which photos should be copied to the cloud.}  

Another example is a smartphone that is programmed to only show the user certain WiFi networks, restricting user choice, or a smartphone that filters out certain recommendations from applications - which can happen without the user's knowledge. 
In general, people would like to maintain choices and freedoms  in the presence of automation - this is also discussed  in the context of automated vehicles later.

In relation to location-based services, or more generally, context-aware mobile services, control and trust are concerns~\cite{Abbas2015UsingAS}. Someone might willingly give away location or contextual information in order to use particular services (an outcome of  a privacy-utility trade-off), assuming s/he trusts the service; the user still retains the choice of opting in, or not, and opting out anytime during the use of the service. Tracking a child for safety can be viewed as somewhat intruding on his/her privacy, but might be insisted on by the parent. As mentioned in~\cite{Abbas2015UsingAS}, in general, a wide range of considerations is required to judge if such context-aware services are ethical or not ethical, including rules and norms, benefits and harms, concerns of people, governing bodies, and cultural values.

\section{Towards a Multi-Pronged Approach}

How one can  build IoT devices that will behave ethically is still a current area of research.
This section reviews a {  range of ideas which have been proposed and applied } to ameliorate the situation, including  how to program ethical behaviour in devices, algorithmic transparency for accountability, algorithmic social contracts and crowdsourcing ethics solutions, enveloping IoT systems, and devising code of ethics for developers. { Then, it is argued that, as each idea  has its own merits and usefulness towards addressing ethical concerns, a multi-pronged approach can be useful.}
 
\subsection{Programming Ethical Behaviour}
We review a range of techniques which have been explored for programming ethical behaviour: rule-based programming and  learning, game-theoretic calculations, ethical settings and ethical-by-design.

\subsubsection{Rule-Based Ethics}
If we want robots or an algorithm to be ethical and transparent, perhaps
one way is to encode the required behaviour explicitly in rules or to create algorithms to allow devices to calculate the ethical actions. 
Rules have the advantage that they 
can be human understandable, and they can represent a range of   ethical behaviours
required of the machine.
 Foundational ideas of ethics such as consequentialism and deontology often underpin the approaches taken.
 {The general idea is that a device whose behaviour abides by these rules is then considered ethical.}

The work in~\cite{10.2307/26002215} 
describes a vision of robots and an example of coding rules of behaviour
for robots.
{\em EthEL}~\cite{DBLP:conf/aaaifs/AndersonA08}
is an example of a robot that provides reminders about medication. There
are issues of when to notify the (human) carer/overseer when the patient
does not take medication. It would be good for the patient to be
respected and have a degree of autonomy to delay or not take medication,
but an issue arises when, the medication, if not taken, leads to a
life-threatening situation - the issue is when the robot should persist
in reminding and inform the overseer, or when it does not, respecting
the autonomy of the patient.

A machine learning algorithm based on inductive logic was used to learn
a general ethical rule about what to do based on particular training cases given
to the algorithm: ``{\em a health care robot should challenge a patient's decision--- violating
the patient's autonomy---whenever doing otherwise would fail to prevent
harm or severely violate the duty of promoting patient welfare.''}
In 2008, this was believed to be the first robot governed by an
ethical principle, where the principle was learned.

{ 
The work in~\cite{8345565} proposes a framework for building ethical agents and the use of norms to characterise interactions among autonomous agents, with three types of norms, namely commitments, authorizations and prohibitions. Such norms can be used by agents needing to behave ethically.  Such multiagent modelling  maps well to decentralized IoT systems allowing the placing of decentralised intelligence and algorithmic behaviour within the IoT~\cite{7980111}. 
}

Ethical questions can be a lot more complex, in general - it would be
hard to encode in rules every conceivable situation where a robot should
persist with its reminders, even when the patient rejects it.
It remains an open research area as to what extent such rules can be
coded up by a programmer, or learnt (via some machine learning
algorithm), for machines in a diverse range of situations in other
applications.

 Another example is the work of  \cite{Gerdes2016} which  
 outlines programming ethical behaviour for autonomous vehicles by mapping ethical considerations into costs and constraints used in automated control algorithms.
Deontological rules are  viewed as constraints in an optimal control problem of  minimising costs, for example, in the case of deciding actions to minimise damages in an incident.
%

From the above  examples, the general overarching rule that saving human life takes priority, over conforming to traffic laws and following a person's (perhaps under-informed) decision.
{However, it is  generally difficult to ensure that a vehicle would abide by these rules - and generally difficult for automated vehicles to assess situations accurately to know which rule applies. Also,
its  software would need to be tested to follow such principles, or testing be done by a certification authority though requiring tremendous resources.}

{
In ~\cite{DBLP:journals/corr/abs-1708-06374}, a formal model of safe and scalable self-driving cars is presented where a set of behavioural rules are specified which could be followed by cars to help ensure safety. 
A rule-based approach could work for specific IoT applications where the rules are identifiable and can be easily encoded.

However, in general, a difficulty is how one could comprehensively determine what the rules are for specific applications, apart from expert opinion. This  raises the question of who decides what is ethical and what is not, and whether users could trust the developers who engineered the IoT systems on what is ethical behaviour.
Apart from experts encodings rules, an alternative approach proposed by MIT researchers is to crowdsource human perspectives on moral decisions, 
as experimented with by the Moral Machine for autonomous vehicles, with interesting results including cross-cultural ethical variation~\cite{nature}.\footnote{http://moralmachine.mit.edu/}
}

System architectures for building machines capable of moral reasoning remains a research area~\cite{Wallach:2010:MMT:1875303,anderson-anderson-2011}.\footnote{Also https://www.nature.com/news/machine-ethics-the-robot-s-dilemma-1.17881\#auth-1} Recent work has proposed rule-based models to encode ethical behaviour that can be computationally verified~\cite{DENNIS20161}, and in contrast to verification approaches, an internal simulation of actions by a machine in order to predict consequences is proposed in~\cite{VANDERELST201856}.

\subsubsection{Game-Theoretic Calculation of Ethics}
Game-theoretic approaches have also been proposed for autonomous vehicles to calculate  ethical decisions, e.g., using Rawlsian principles in contrast to utilitarian approaches~\cite{Leben:2017:RAA:3095624.3095632}. The idea is to determine, given the behaviour of the party, the best outcome.
A difficulty is deciding whether Rawlsian or a Utilitarian calculation should be employed, or even other schemes - the Rawlsian approach aims to maximise utility for the worst case (maximin approach) while the Utilitarian approach aims to maximise utility.  It is also difficult to assign appropriate numerical values for the utilities of actions (e.g., why would hitting a pedestrian have a value of -1 while injuring a pedestrian is given a value of -0.5?).

\subsubsection{Ethics Settings}
 Another category of work focuses on getting user input in `programming' the ethical behaviour of devices, in particular, for autonomous cars. The notion of {\em ethics settings} or the ``ethical knob'' was proposed by~\cite{Contissa2017}, to allow passengers of autonomous vehicles to make choices about ethical dilemmas, rather than have the reasoning hard-coded by  manufacturers. For a vehicle needing to prioritise between the safety of the passengers and of pedestrians in road situations, there are three modes, namely {\em altruistic}, {\em egoistic} and {\em impartial}, corresponding to the preferences for the safety of the pedestrian, the safety of the passenger and the safety of both, and the passenger can choose a mode somewhere in between, among the three. The idea of the ethics settings is advocated by~\cite{Gogoll2017}, which also answers the question of what settings people should use - each person choosing the selfish ethics settings might make society worst off overall, while everyone, if this can be mandated, choosing the settings that minimise harm, even if altruistic, would make society better off.

\subsubsection{Ethical by Design}

In~\cite{Baldini2016}, an approach is to allow designers of IoT systems to configure via a set of available policy templates, which reduces the complexity of the software engineering of IoT systems, where multiple policies are relevant, e.g., a policy on storage of data, a policy on how data can be shared  or a policy on ethical actions. A set of policies can be chosen by the user or the developer (in view of the user) that is tailored to the user's capabilities and context.
A framework for dynamic IoT policy management has been given in~\cite{8027013}.

{
While this review does not focus on challenges of IoT data privacy specifically, the review in~\cite{seliem2018} noted that addressing IoT data privacy challenge involves designing and building in data access control and sharing mechanisms into IoT devices, e.g., building in authentication and key establishment mechanisms in IoT devices, computing on the edge to address privacy concerns,  mechanisms to mask shared personal   data, tools to support user access rules for data handling, and tools for digital forgetting and data summarization.  In summary, one can reduce user privacy leakage and risks of IoT devices mishandling IoT data via a combination of these mechanisms. 

When such mechanisms are known and existing, and as more of such mechanisms are developed, then according to~\cite{Baldini2016}, ethically designed   IoT products (including devices and applications) are those  ``designed and deployed to empower users in controlling and protecting their personal data and any other information.''  The idea of  the ``ethical knob'' also seeks to put more control into the hands of users, beyond data handling. Hence, to program in ethical behaviour is not only programming IoT devices that take action based on ethical considerations, but also providing users appropriate control over device behaviour even as it has delegated authority to act autonomously.
}

   \subsection{Enveloping IoT Systems} 
 The concept of ``enveloping'' was first introduced in~\cite{Floridi2019} in regard to providing boundaries within which today's AI systems can work effectively. A distinction is made between complexity of a task, in relation to how much computational resources it requires, and the difficulty of the task, relating to the physical manipulation skills it requires, e.g., the gross or fine motor skills (robotic or human) required to perform tasks such as dish washing with hands, painting with a brush, tying shoe-laces, typing, using a tool, running up the stairs, playing an instrument, or helping somewhat disabled walk or get up. Examples, taken from the paper, of {\em envelopes} for devices include, for industrial robotics, ``the three-dimensional space that defines the boundaries within which a robot can work successfully is defined as the robot's envelope", and the waterproof box of a dishwasher, and Amazon's robotic shelves and warehouse for its warehouse robots. It is noted that ``driverless cars will become a commodity the day we can successfully
envelop the environment around them.'' A computer chess program can be very successful within the constraints of the rules of chess. Indeed, the idea of dedicated lanes or areas for  automated vehicles   can be viewed as a type of envelope for such vehicles.  Hence, enveloping is a powerful idea for successful AI systems.
 
While it might not always be possible to envelop IoT systems, consider a generalized view of enveloping that is not just physical, but also cyber-physical, comprising the situation spaces (physical boundaries and cyber boundaries)  in which a device functions. Such enveloping can  help in addressing ethical issues, by reducing the complexity of the environment in which such IoT devices or robots operate, reducing  the chance for unintended situations, allowing comprehensive rules to be devised in a more constrained operating environment,  helping to manage human expectations (e.g., humans generally get out of the way of trains, trams and vehicles on the road), and enabling clear definition of the context of operation, e.g.,  algorithmic bias is then not unexpected if the context of the development of the algorithm is known, such as the training dataset used,  and the Internet  environment or ``cyber-envelope'' in which the device operates, including where data is stored and shared is explicitly co-defined, by IoT device manufacturers and users. Also, as another example, a pill-taking reminder system works within its known envelope so that unexpected  behaviours when working beyond its envelope could be expected. However, enveloping can prove restrictive in the IoT, and successful enveloping to help deal with ethical IoT issues is still to be proven.

\subsection{Whitebox Algorithms}

As noted earlier, algorithms might be used to make decisions, that
affect people in a significant way, e.g., criminal cases and whether someone
should be released from prison, to whether someone is diagnosed with a
particular disease. Also, certain groups of people may feel unfair if an
algorithm did not work as well for them as it did for someone on account
of his/her skin colour or accent.

How can one deal with algorithmic bias? Two areas of research to
address this problem are noted: algorithmic transparency and detecting algorithmic bias.

\subsubsection{Transparency}\label{transparency}

There are at least two aspects of transparency for IoT devices - the data traffic going into and out of such devices,
and the inner workings of such devices.

For example,
the TLS-RaR approach~\cite{Wilson:2017:TBV:3081333.3081342} allows device owners (or consumer watchdogs
and researchers) to audit the traffic of their IoT
devices.  Affordable in-home devices called {\em Auditors} can be added to the local network to observe 
network data for IoT devices using  Transport Layer Security (TLS) connections. However, some devices might use steganography to hide data in traffic, or users might still miss some data sent out by a malicious device.

Apart from monitoring the traffic of IoT devices,
there are many who argue that algorithms that make important
decisions should be a ``white box'' rather than a ``black box'' so that
people can scrutinise and understand how the algorithms make decisions,
and judge the algorithms that judge us. 
{This is  also a current emphasis for the explanability for Artificial Intelligence (AI) idea.\footnote{https://en.wikipedia.org/wiki/Explainable\_Artificial\_Intelligence}
This can become an increasingly important feature for IoT devices that take action autonomously - users need to know why, for example, heating has been reduced in a smart home recently, forgetting that a target energy expenditure was set earlier in the month.}

{For widely used systems and devices where many people could be affected, 
 transparency enables social accountability. 
IoT devices in public spaces, deployed by town council, should work according to public expectations.
IoT public street lights that  systematically only lights up certain segments of a road for particular shops, and not for other shops, 
can be seen to be bias - or at least, must be in error and be subject to scrutiny. }

{Consider IoT devices whose purpose is to provide information to people, or devices that filter and provide information to other devices;
transparency in such devices enable people to understand (at least in part) why certain information is shown to them, or understand their behaviour.}
Example,   Facebook has been rather open in how its
newsfeed algorithm works\fnurl{https://blog.bufferapp.com/facebook-news-feed-algorithm}{.} 
 It can be seen that by
being open about how the algorithm works, Facebook provides, to an
extent, a form of social
accountability. 

Another way an algorithm could ``expose'' its workings is to output logs
of its behaviour over time. For example, in the case of autonomous
vehicles, if an accident happens between a human-driven car and an
autonomous vehicle, one should be able to inspect logs to trace back to
what happened and decide if the company should be held accountable or
the human driver. This is similar to flight recorders in commercial
airplanes. 
{As another example of auditing, the Ditio~\cite{Mirzamohammadi:2017:DTA:3131672.3131688}  system is an approach for auditing sensor activities, logging activities that can be later inspected by an auditor and checking for compliance with policies. An example is given where the camera on a Nexus 5 smartphone is audited to ensure that 
 it has not been used for recording during a meeting. 
}

However, there are concerns with logging and whitebox views of
algorithms. For example, intellectual property might be a concern when
the workings of an algorithm is transparent, or when data used in
training a machine learning algorithm is exposed. Care must be taken in
how algorithms are made transparent. Another issue is that, often, with neural network learning
algorithms, the actual rules learnt for classification and decision-making  are also not
explicitly represented (and are simply encoded in the parameters of the
neural network).
Also, what type of data or behaviour should be logged and how can they be managed
remains open issues and are application-dependent.

The white box algorithm approach can be employed to expose
algorithmic bias when present or to allow human judgement on algorithmic
decisions, but the workings of a complex algorithm is not easily legible
or understandable in every situation.

Algorithms and systems may need to be \emph{transparent by design} - a software engineering challenge.
The paper
on ``Algorthmic Accountability'' by the World Wide Web Foundation\footnote{http://webfoundation.org/docs/2017/07/Algorithms\_Report\_WF.pdf}  calls
for explainability and auditability of software systems, in particular
those based on machine learning, to encourage algorithmic accountability
and fairness. The Association of Computing Machinery (ACM), a major
computing association based in USA, calls for \emph{algorithmic
transparency and accountability} in a statement.\footnote{https://www.acm.org/binaries/content/assets/public-policy/2017\_usacm\_statement\_algorithms.pdf}

Getting algorithms or systems to explain their own actions and audit
their own execution have become current research areas, as suggested by
this workshop on Data and
Algorithmic Transparency.\footnote{http://datworkshop.org/\#tab\_program}

{
Also, using  open source software  has been argued as an approach to achieve transparency, e.g, of AI algorithms.\footnote{https://www.linuxjournal.com/content/what-does-ethical-ai-mean-open-source} However, commercial interests might hinder free and open software.
 }
 
{ 
In summary, for transparency and accountability,  as noted in \cite{8423131}, IoT systems
from the technical point of view can provide {\em control} - allowing users to control what happens,
and {\em auditing} - enabling what has happened or why something is happening to be recorded. 
IoT systems also need to allow users to understand (and perhaps control) what data they collect and what they do with that data,
 to allow users to understand (and perhaps configure) their motivations,\footnote{http://iot.stanford.edu/retreat15/sitp15-transparency-no.pdf} 
 and to see (and perhaps change), in a non-technical way, how they work.
}

\subsubsection{Detecting Algorithmic Bias}\label{detecting-algorithmic-bias}

 People might stumble upon such bias when using some devices, but sometimes it may be a lot more subtle (e.g., considering a news
feed where we may not realize or miss what we are not expecting).
Researchers have looked at how to detect algorithmic bias using
systematic testing based on statistical methods, such as in a technique called  {\em transparent model
distillation}~\cite{DBLP:journals/corr/abs-1710-06169} which we will not look at in-depth here.

\subsection{Blackbox Validation of Algorithmic Behaviour}
There could be situations where whiteboxing algorithms is not possible due to commercial reasons, and generating explanations from certain (e.g., deep learning) algorithms is still a challenge. 
It is well articulated in~\cite{mb} that 
\begin{quote} {\em
``the study of machine behaviour will often require experimental intervention to study human–machine interactions in real-world settings''
}
\end{quote}
Software testing is well studied and practised. Experimental evaluation of algorithmic behaviour to verify certain properties or capabilities might be employed, though testing device behaviour in all circumstances and environments can be challenging, especially if it connects to other devices and if there are  flow-on consequences in the physical world, and considering the complexity of a device. A notion of the Turing test has been proposed for autonomous vehicles.\footnote{For example, see https://news.itu.int/a-driving-license-for-autonomous-vehicles-towards-a-turing-test-for-ai-on-our-roads/}

Where the range of possible situations and interactions with the environment are complex, apart from real-world testing, simulation-based testing and validation can be an economical solution, as noted in~\cite{10.1007/978-3-658-16988-6_106}, as an example. Also, software updates are expected to occur with IoT devices and validation might then need to be redone, changes localised to particular modules, and the impact of changes on other modules assessed  - the work in~\cite{Ebert2019,8802868} noted that testing of autonomous vehicles and autonomous systems requires such {\em cognitive testing} as it is called.

The design of the human-device interface is also a consideration if users are to exercise choice and freedom, they would need to understand the functions of the device and how to interact with it. The interface should not be too complex so that users lose comprehension yet should make available adequate choices and options - this is indeed a challenging task for a complex device. For example, for automated vehicle Human-machine interfaces (HMIs), using heuristic evaluation is one approach~\cite{NAUJOKS2019121}, where a set of criteria is used to judge the HMI.

Validation requires criteria to validate against - a safety standards approach for fully autonomous vehicles has been proposed in \cite{10.1007/978-3-030-26250-1_26}. Similar standards of algorithmic behaviour might be devised for other types of IoT devices, e.g., for delivery robots on walkways, or robots in aged care homes.
 
\subsection{Algorithmic Social Contracts}

Going beyond the simple white box concept for algorithms, the work in~\cite{Rahwan2017} proposed a conceptual framework for
regulating AI and algorithmic systems.

The idea is to create social contracts between stakeholders that can
hold an algorithmic system (and its developers) accountable and allow
voting and negotiation on trade-offs (e.g., should a feature that
increases pedestrian safety in autonomous vehicles but decrease
passenger safety be implemented? Or should a feature of a system that
decreases data privacy but increases public safety be deployed?). The
idea is to: {\em
`to build institutions and tools that put the society in-the-loop of
algorithmic systems, and allows us to program, debug, and monitor the
algorithmic social contract between humans and governance algorithms.'}

What is proposed is for tools to be developed that can take technical
aspects of algorithms and present that to the general public so that
they can be engaged in influencing the effect and behaviour of the
algorithms - effectively {\em crowdsourcing ethics}, an approach used elsewhere~\cite{Lieberman:2013:CEP:2468356.2468481}. The general approach of combining machine-learned representations and human perspectives has also been called {\em lensing}.\footnote{https://www.media.mit.edu/videos/2017-05-18-karthik-dinakar/}

Tim O'Reilly's 2016 book  {\em ``Beyond Transparency: Open Data and the Future of Civic Innovation}'' proposes the idea of
\fnurl{http://beyondtransparency.org/chapters/part-5/open-data-and-algorithmic-regulation/}{algorithmic
regulation,} where algorithmic regulation is successful when there are:
``{\em (1)  a deep understanding of the desired outcome, (2)
  real-time measurement to determine if that outcome is being achieved,
(3) algorithms (i.e.~a set of rules) that make adjustments based on new
  data, and (4)
  periodic, deeper analysis of whether the algorithms themselves are
  correct and performing as expected.''}
The actual processes to achieve the above is a still an unresolved
socio-technical challenge, in itself an area of  research.

\subsection{Code of Ethics and Guidelines for IoT Developers}

Rather than building ethical behaviour into machines, ethical guidelines are also useful for the developers of the technology.  
There are codes of ethics for robotics engineers,\footnote{https://web.wpi.edu/Pubs/E-project/Available/E-project-030410-172744/unrestricted/\\A\_Code\_of\_Ethics\_for\_Robotics\_Engineers.pdf} and more recently, the
{\em Asilomar Principles}\label{asilomar-principles} for AI research.
These principles were developed in conjunction with the 2017 Asilomar
conference and relates to ethics in AI R\&D.\footnote{https://futureoflife.org/ai-principles/}
The principles   cover safety, transparency, privacy, incorporating human values and maintaining human control. 
The notion of how to imbue algorithms and
systems with human values is a
\fnurl{http://www.valuesincomputing.org/}{recent research topic.} The
above appears to provide a morally sound path for AI R\&D and AI
applications, and for IoT devices with AI capabilities.
Code of ethics for IoT is also being discussed.\footnote{See the interview at https://www.theatlantic.com/technology/archive/\-2017/05/internet-of-things-ethics/524802/,\\and EU discussions at\\http://ec.europa.eu/transparency/regexpert/index.cfm?\-do=groupDetail.groupDetailDoc\&id=7607\&no=4}  An IoT design manifesto\footnote{https://www.iotmanifesto.com/} 
presents a range of general design principles for IoT developers.  { The IoT Alliance Australia has provided security guidelines.\footnote{https://www.iot.org.au/wp/wp-content/uploads/2016/12/IoTAA-Security-Guideline-V1.2.pdf} }

{The German Federal Minister of
Transport and Digital Infrastructure appointed a national ethics
committee for automated and connected driving which presented a code of ethics for automated and connected driving~\cite{Luetge2017-LUETGE}. The ethical guidelines highlights a range of principles including ``{ \em ...Technological development obeys the principle of
personal autonomy, which means that individuals enjoy freedom of action for which
they themselves are responsible}'', i.e., personal autonomy is a key principle for ethical technology. Also, autonomous cars can improve the mobility of the disabled, and so, has ethical benefits. Another guideline stresses that official licensing and monitoring are required for automated driving systems - which may be a direction required for robotics autonomous things in public, from drones to delivery robots. A controversially debated guideline involves {\em ``..General programming to reduce the
number of personal injuries may be justifiable}'' even at the cost of harm to some others, which somewhat adopts a utilitarian view of ethics which may not be agreeable to all.
Another guideline on accountability: ``{ \em that manufacturers or
operators are obliged to continuously optimize their systems and also to observe
systems they have already delivered and to improve them where this is technologically
possible and reasonable}'' applied to automated vehicles, but suggests possible implications on ethical IoT, in general, raising the question of the responsibility for maintenance and continual upgrades by manufacturers or operators for IoT devices post-deployment. Privacy-by-design is a principle suggested for data protection for connected driving.}

{
It remains to be seen how ethical principles can be software engineered into
future systems and whether certification requirements by law is possible, especially in relation to data handling by IoT devices.\footnote{https://www.researchgate.net/publication/322628457\\\_The\_Legal\_Challenges\_of\_Internet\_of\_Things}   }

{
The CTIA Cybersecurity Certification Test Plan for IoT devices\footnote{https://api.ctia.org/wp-content/uploads/2018/08/CTIA-IoT-Cybersecurity-Certification-Test-Plan-V1\_0.pdf} aims to  define tests to be conducted on IoT devices for them to be certified in terms of three levels of security features built-in. Other standards and
guidelines for IoT data privacy and device security are also being
\fnurl{https://www.schneier.com/blog/archives/2017/02/security\_and\_pr.html}{proposed and developed.}}

{
A comprehensive framework to help researchers  identify and take into account  challenges in ethics and law when researching IoT technologies, or more generally, heterogeneous systems, is given in~\cite{8379713}. Review of research projects by an ethics review board, consideration of national/regulatory guidelines and regulatory frameworks, and wider community engagement are among the suggested actions.
}

 {
 On a  more philosophical note is the question: what guidelines and strategies (or pro-social thinking) for the addition of each new device to the Internet of Things can 
 encourage favourable evolution of the Internet of Things even as it is being built?  This is a generally challenging issue, especially in a competitive world, but the mechanisms of reciprocity, spatial selection, multilevel selection and kin selection are known to encourage cooperation~\cite{RAND2013413}. Prosocial preferences sometimes do not explain human cooperation~\cite{Burton-Chellew216}, and  the question of how favourable human cooperation  can happen continues to be explored, even from the viewpoint of   models from statistical physics~\cite{PERC20171}.}

 The work in~\cite{STAHL2016152} argues for embedding ethics into the development of robotics in healthcare  via the concept of Responsible Research and Innovation (RRI). The RRI provides a toolkit of questions that helps to identify, understand  and address ethical questions when applied to  this domain.


 \subsection{Summary and Discussion}
 
{ In summary, we make the following observations:}
\begin{itemize}
\item {\em A multi-pronged approach.} Table~I summarises the above discussion detailing the ideas and their main methods with their key advantages and technical challenges. It can be seen that each idea has advantages and challenges, and they could complement each other so that combinations of ideas could be a way forward. { Combining process and artifact strategies would mean taking into account ethical guidelines and practices in the development of IoT devices, and where applicable,  also building functionality into the device which allows the device to behave in an ethical manner (according to an agreed criteria) during operation. Devices can be built to work within the constraints of their enveloping environment, with user-informed limitations and clear expectations in terms of applicability, configurability, and behaviour.  Developers could encode rules for ethical behaviour, but only after engagement and consultation with the community and stakeholders on what rules are relevant, based on a transparent  and open process (e.g., consultative processes, technology trials, crowdsourcing viewpoints or online workshops). White or gray boxed devices could allow end-user intelligibility, consent and configurability, so that users retain a desired  degree of control.}
{  Individual IoT devices should be secured against certain cyber-attacks, and then the data they collect should be handled in a way that is intelligible and configurable by the user, according to best practice standards, and when they take action, it should be in agreement with acceptable social norms, and auditable.}

{ 
\item {\em Context is key.} Many people could be involved in an IoT ecosystem, including developers, IoT device retailers, IoT system administrators, IoT system maintainers, end-users, local community and society at large. Society and communities can be affected by the deployment of such IoT systems in public, e.g., autonomous vehicles and robots in public, and so, the broader context of deployment needs to be considered. 

Moreover, what is considered ethical behaviour might depend on the context of operation  and the application - a device's action might be be considered ethical in one context but unethical in another, as also noted in~\cite{Abbas2015UsingAS} with regards to the use of location-based services. Broader contexts of operations include local culture,  norms, or application domain (e.g., IoT in health, transport, or finance would have different rules for ethical behaviour); hence, it would require multiple levels of norms and ethical rules to guide the design and development of IoT devices and ecosystems: a basic ethical standard could apply (e.g., basic security built into devices,  basic user-definable data handling options, and basic action tracking), and then  additional configurable  options for context-specific ethical behaviour added.

\item {\em Ethical considerations with autonomy.} Guidelines for developers could encourage thinking through the following, and what is built into an artifact to achieve ethical algorithmic behaviour could incorporate features that take into account at least the following:
\begin{itemize}
\item security of data and physical security as impacted on by device actions,
\item privacy of user data and device actions that impinge on privacy,
\item consequences of over-reliance or human attachment to IoT devices,
\item algorithmic bias and design bias, and fairness of device actions,
\item the possible need to engage not just end-users but anyone affected by the IoT deployment, e.g., via crowdsourcing viewpoints pre-development, and obtaining  feedback from users and society at large post-deployment,
\item user choice and freedom retained,  including allowing user adjustments to ethical behaviour (e.g., opt in and out, adequate range of options, and designing devices with ethical settings),
\item end-user experience including user intelligibility, scrutability and explainability when needed, usability not just for certain groups of people,  user control over data management and device behaviour, and appropriate manual overrides\footnote{https://cacm.acm.org/blogs/blog-cacm/238899-the-autocracy-of-autonomous-systems/fulltext},
\item accountability for device actions, including legal and moral responsibilities, and support for traceability of actions, 
\item  implications and possible unintended effects of cooperation among devices, e.g., where physical actions from multiple devices could mutually interfere, and the extent of data sharing during communications,
\item deployment for long-term use (if applicable) and updatability, arising from security updates,  improvements from feedback, adapting to changing human needs, policy changes,   and
\item  ethical consequences of  autonomous action in IoT deployments (from physical movements to driving in certain ways).
\end{itemize}
}
\end{itemize}

\begin{table}
\scalebox{0.68}{
    \begin{tabular}{| l | l |}
    \hline  &  
      \begin{tabular}{ p{3cm} | p{3cm} | p{3cm} | p{4cm}| p{4cm}}
{\bf Idea} & {\bf Methods}  & {\bf Key Advantages}  & {\bf Key Challenges}  & {\bf  Selected Related Work}  
      \end{tabular} \\ \hline \hline
    \cellcolor{cyan}    \rotatebox{90}{\bf \color{white} Build Behaviour into Artifact \& Validate} & 
\begin{tabular}{ p{3cm} | p{3cm} | p{3cm} | p{4cm}| p{4cm} }
\rowcolor{white} Designing and Programming \newline Ethical Behaviour &  rule-based, \newline game-theoretic calculations, \newline ethics settings, \newline ethical design templates &   algorithmic or declarative representation  of ethical behaviour,  user control explicitly considered in artifact design  &  difficult for a set of rules to be complete, { data used in development (e.g., to train Machine Learning models used in IoT devices) might be inadequate}, hard for situations to be quantified,  raises the question of who decides what is ethical & \cite{10.2307/26002215,DBLP:conf/aaaifs/AndersonA08,8345565,7980111,DBLP:journals/corr/abs-1708-06374,Leben:2017:RAA:3095624.3095632,Contissa2017,Gogoll2017,Baldini2016,8027013}  \\ \hline
\rowcolor{white}  Enveloping &  setting physical / cyber-physical boundaries of operation &  reducing complexity of operating environments, \newline sets expectations in behaviour   and contexts of trustworthy operation &  may be hard to create suitable envelopes that do not hinder functioning of IoT systems & \cite{Floridi2019} (though originally proposed to achieve better AI systems)   \\ \hline
\rowcolor{white} Whitebox\newline Algorithms &  improve transparency, \newline detect algorithmic bias &  greater traceability and accountability ({\em possibly allow engagement with non-developers}) &  transparency does not equate understandability,  scrutability does not equate user control,  & \cite{ Wilson:2017:TBV:3081333.3081342,Mirzamohammadi:2017:DTA:3131672.3131688,8423131,DBLP:journals/corr/abs-1710-06169} \\ \hline
\rowcolor{white} Blackbox\newline Validation &  cognitive testing, \newline simulation \newline heuristic evaluation  & where white-boxing is difficult, \newline basis for certification &  difficult to consider all cases and situations & \cite{10.1007/978-3-658-16988-6_106,8802868,NAUJOKS2019121} \\ \hline
\rowcolor{white} Algorithmic  \newline Social Contracts &  crowdsourcing ethics, \newline processes for algorithmic regulation &  wider engagement ({\em possibly with non-developers}) &  complex, may be hard to create suitable  efficient processes or  to obtain adequate participation & \cite{Rahwan2017,Lieberman:2013:CEP:2468356.2468481} 
\end{tabular} \\  \hline
    \cellcolor{blue}    \rotatebox{90}{\bf \color{white} Guide Developers } & 
\begin{tabular}{p{3cm} | p{3cm} | p{3cm} | p{4cm}| p{4cm}}   
\rowcolor{white} Code of Ethics and Guidelines \newline for IoT  Developers & formal guidelines, \newline regulations, \newline community best practice for developers  & highlights ethical considerations in development & application or domain specific considerations required  &  German ethics code for automated and connected driving~\cite{Luetge2017-LUETGE}, IoT data privacy  guidelines and regulations~\cite{8453643}, \newline (also, Code of ethics for robotics engineers, \newline Asilomar Principles, IoT design manifesto,  IoT Alliance Australia Security Guideline,  \newline design-for-responsibility), RRI~\cite{STAHL2016152} \\ \hline
\end{tabular}
\\ \hline
 \end{tabular}
 }
\caption{Summary of Ideas to Achieve Ethical algorithmBehaviour in the IoT with key advantages and challenges.}
\end{table}

\section{Conclusion and Future Work}

This paper has reviewed a range of ethical concerns with IoT, including concerns that arises when IoT technology is combined with robotics and AI technology (including machine learning) and autonomous vehicles.
The concerns include informational security, data privacy,  moral dilemmas, roboethics, algorithmic bias when algorithms are used for decision-making and control of IoT devices, as well as risks in cooperative IoT.

The paper has also reviewed approaches that have been proposed to address these ethical concerns, including
\begin{itemize}
\item  programming approaches to add ethical behaviour to devices, including adding moral reasoning capabilities to machines, and configuring devices with  user ethics preferences,
\item detection and prevention of algorithmic bias, via accountability models and transparency,
\item behaviour-based validation techniques,
\item the notion of algorithmic social contracts, and  crowdsourcing solutions to ethical issues, 
\item the idea of enveloping systems, and
\item  developing guidelines and proposals for regulations,  and codes of ethics, to encourage ethical developers and ethical development of IoT devices, and requiring security and privacy measures in such devices. {Suitable data privacy laws in the IoT context, secure-by-design, privacy-by-design, ethical-by-design and design-for-responsibility principles will also be needed.}
\end{itemize}

{ A multi-pronged approach could be explored to achieve ethical IoT behaviour in a specific context. More research is required to explore combined approaches, and to create a framework of multiple levels of ethical rules and guidelines that could cater for the context-specific nature of what constitutes ethical behaviour.}

This paper has not considered in detail legislation and the law involving robots and AI, approaches of which could be considered for intelligent IoT systems, which are addressed in depth elsewhere~\cite{DBLP:books/daglib/0032642}.
{ Also, the notion of IoT policing has not been discussed}, in the sense of run-time monitoring of devices to detect misbehaving devices, perhaps with the use of sentinel devices, as well as policy enforcement, penalties imposed  on anti-social IoT devices (e.g., game-theoretic grim-trigger type strategies, and other types of sanctions for autonomous systems~\cite{journals/ker/NardinBAKSS16}).
Social equity and social inequality are two concerns of the social ethics of the Internet of Things which have been discussed elsewhere~\cite{8390757} but not detailed here. {
Sustainability of IoT deployments~\cite{stead}\footnote{https://www.computerworld.com.au/article/561064/hidden-environmental-cost-internet-things/} and the use of IoT for sustainability~\cite{BIBRI2018230}  have not been  extensively discussed here, which have socio-ethical implications.  }

{The challenge of building ethical things in the IoT that act autonomously yet ethically will also benefit from on-going research in building ethics into AI decision-making as reviewed in~\cite{ijcai2018-779}, which  includes individual ethical decision frameworks,  collective ethical decision frameworks, ethics in human-AI interactions and systems to explore ethical dilemmas.}

Outstanding socio-technical challenges remain if IoT devices are to  behave ethically and be used ethically, for IoT developers and IoT users. Ethical considerations would need to be factored into future IoT software and hardware development processes, according to upcoming  certification practices, ethics policies, and regulatory frameworks,  which are still to be developed. Particular domains or contexts would require domain-specific guidelines and ethical considerations. 

While we have addressed mainly ethical behaviour for IoT device operations and the algorithms therein, there are ethical issues  concerning the post-deployment and maintenance of IoT devices, where retailers or manufacturers could  take responsibility.

\bibliographystyle{plain}
\bibliography{ethics-iot}

 \end{document}